\documentclass[twocolumn,trackchanges]{aastex7}
\usepackage{graphicx}
\usepackage{hyperref}
\usepackage{color}
\usepackage{dirtytalk}
\usepackage{multirow}

\received{May 14, 2025}
\revised{July 10, 2025}
\accepted{July 16, 2025}

\begin{document}

\title{JVLA and VLBA study of the merging cool core CHIPS~1911+4455 at z$\sim$0.5 \\Radio emission from an infant AGN and from a rapidly star-forming BCG }
\shorttitle{JVLA and VLBA study of the merging cool core CHIPS~1911+4455 at z$\sim$0.5}

\author[orcid=0000-0001-5338-4472]{Francesco Ubertosi}
\affiliation{Dipartimento di Fisica e Astronomia, Università di Bologna, via Gobetti 93/2, I-40129 Bologna, Italy}
\affiliation{Istituto Nazionale di Astrofisica - Istituto di Radioastronomia (IRA), via Gobetti 101, I-40129 Bologna, Italy}
\email[show]{francesco.ubertosi2@unibo.it}  

\author[orcid=0000-0002-0843-3009]{Myriam Gitti}
\affiliation{Dipartimento di Fisica e Astronomia, Università di Bologna, via Gobetti 93/2, I-40129 Bologna, Italy}
\affiliation{Istituto Nazionale di Astrofisica - Istituto di Radioastronomia (IRA), via Gobetti 101, I-40129 Bologna, Italy}
\email{myriam.gitti@unibo.it}

\author[orcid=0000-0002-8341-342X]{Pasquale Temi}
\affiliation{NASA Ames Research Center, MS 245-6, Moffett Field, CA 94035-1000, USA}
\email{pasquale.temi@nasa.gov}  

\author[orcid=0000-0002-5671-6900]{Ewan O'Sullivan}
\affiliation{Center for Astrophysics $|$ Harvard \& Smithsonian, 60 Garden Street, Cambridge, MA 02138, USA}
\email{eosullivan@cfa.harvard.edu}  

\author[orcid=0000-0001-6638-4324]{Valeria Olivares}
\affiliation{Departamento de F\'isica, Universidad de Santiago de Chile, Av. Victor Jara 3659, Santiago 9170124, Chile}
\affiliation{Center for Interdisciplinary Research in Astrophysics and Space Exploration (CIRAS), Universidad de Santiago de Chile, Santiago 9170124, Chile}
\email{valeria.olivares@usach.cl}  

\author[orcid=0000-0002-4962-0740]{Gerrit Schellenberger}
\affiliation{Center for Astrophysics $|$ Harvard \& Smithsonian, 60 Garden Street, Cambridge, MA 02138, USA}
\email{gerrit.schellenberger@cfa.harvard.edu} 

\author[orcid=0000-0001-9807-8479]{Fabrizio Brighenti}
\affiliation{Dipartimento di Fisica e Astronomia, Università di Bologna, via Gobetti 93/2, I-40129 Bologna, Italy}
\affiliation{University of California Observatories/Lick Observatory, Department of Astronomy and Astrophysics, Santa Cruz, CA 95064, USA}
\email{fabrizio.brighenti@unibo.it}  

\author[orcid=0000-0002-8657-8852]{Marcello Giroletti}
\affiliation{Istituto Nazionale di Astrofisica - Istituto di Radioastronomia (IRA), via Gobetti 101, I-40129 Bologna, Italy}
\email{marcello.giroletti@inaf.it}

\begin{abstract}
Recent studies of galaxy clusters found peculiar cases at the boundary between non-cool core and cool core systems. While unusual, these objects can help us understand the evolution of the most massive clusters. We investigated the role of active galactic nucleus (AGN) feedback in the starburst brightest cluster galaxy (BCG) of the merging cool core cluster CHIPS~1911+4455 (z = 0.485). We conducted new multifrequency (0.3 -- 5 GHz) Very Long Baseline Array (VLBA) and Jansky Very Large Array (JVLA) observations of CHIPS~1911+4455 across a wide range of scales (0.01 to 20 kpc). Our analysis reveals that the AGN in the BCG has recently awakened, showing a compact core with symmetric, $\sim$30 pc long jets in VLBA data. The onset of the AGN may be linked to the enhanced cooling of the hot gas found in a previous study. At larger scales (10~kpc), faint radio whiskers extending to the south show a striking alignment with star-forming knots and are thus interpreted as synchrotron-emitting regions associated with the starburst BCG. The implied radio star formation rate of $100 - 155$~M$_{\odot}$/yr agrees with the optical/infrared one ($140 - 190$~M$_{\odot}$/yr). Our JVLA and VLBA radio study, informed by previous X-ray/optical/millimeter works, indicates that CHIPS~1911+4455 represents a transitional phase in cluster evolution, where the AGN in the central galaxy has just begun to respond to copious hot gas cooling.

\end{abstract}

%

\section{Introduction}
\begin{figure*}[ht!]
    \centering
    \includegraphics[width=\linewidth]{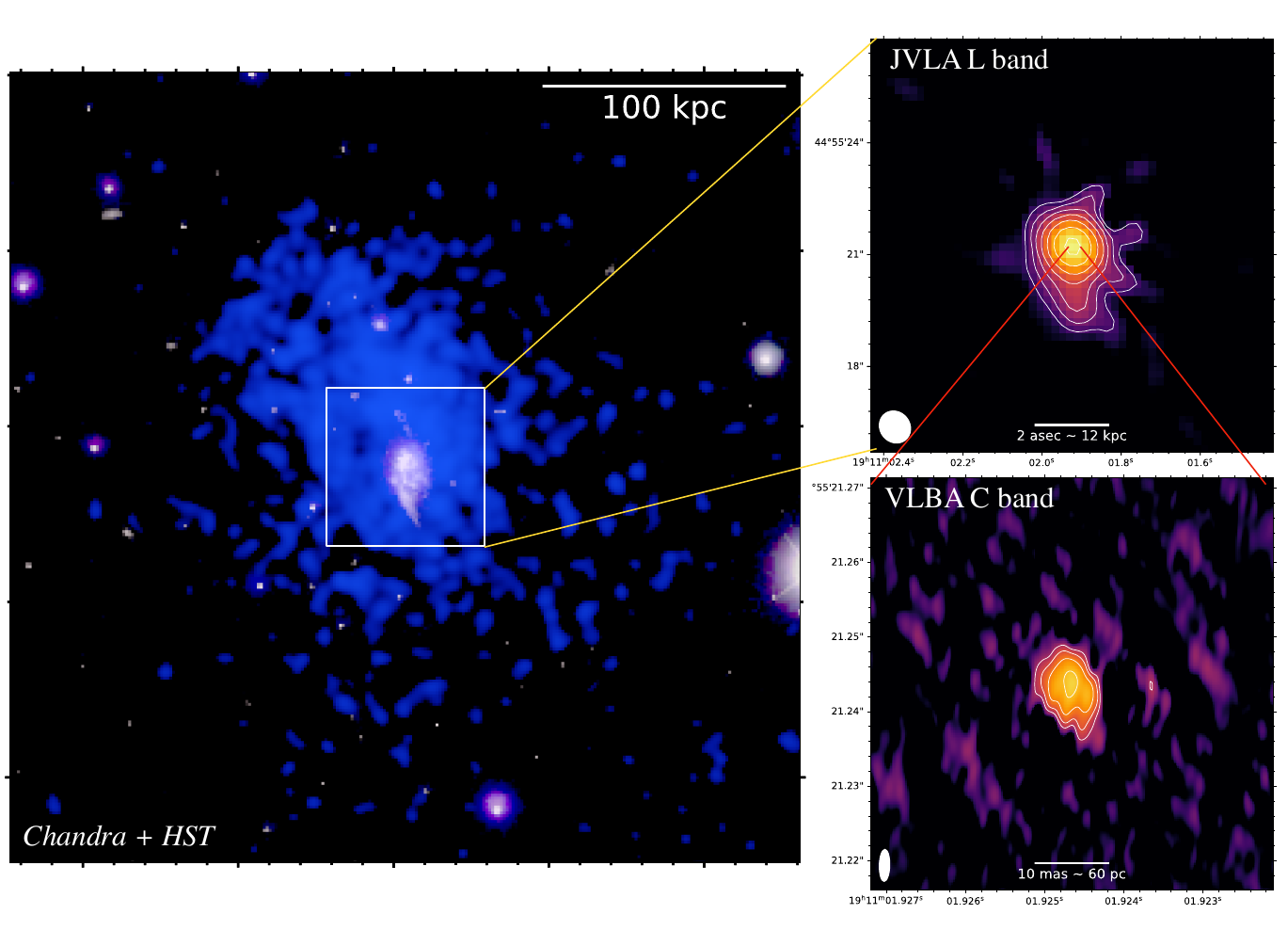}
    \caption{Multi-wavelength view of CHIPS~1911+4455. {\it Left:} Large-scale composite image from archival data (blue from {\it Chandra}, white from HST). The white box shows the extent of the top right panel. {\it Right:} JVLA image at L band (top) and VLBA image at C band (bottom) of the BCG in CHIPS~1911+4455 from the new data presented in this work. Contours are drawn at 5$\times\sigma_{rms}$ and increase by a factor of 2 (see Sec. \ref{sec:data} for details), and the white ellipse represents the beam. }
    \label{fig:multiwav}
\end{figure*}
Galaxy clusters are commonly divided into unrelaxed and relaxed systems (e.g., \citealt{2010A&A...513A..37H,2011A&A...532A.123R,2015ApJ...813L..17R}). In the former, the intracluster medium (ICM) exhibits a highly asymmetric thermal distribution (e.g., \citealt{govoni2004,botteon2018}), due to violent mergers that perturb the gravitational potential (e.g., \citealt{sarazin2002}). The latter are centrally concentrated and show a fairly symmetric temperature structure (e.g., \citealt{2009ApJS..182...12C,2010A&A...513A..37H,andradesantos2017}). In these systems, the efficient cooling of the hot gas determines the emergence of a central cool and dense region, the cool core (see \citealt{molendi2001} for the first occurrence of this term), which surrounds the brightest cluster galaxy (BCG). A key mechanism occurring in cool cores is feedback from the active galactic nucleus (AGN) in the BCG. Radio jets driven by the supermassive black hole (SMBH) deposit energy in the halo, thus regulating cooling of the gas (e.g., for reviews, \citealt{2007ARA&A..45..117M,2012NJPh...14e5023M,2022PhR...973....1D}).
\\\par Despite the fundamental distinction between cool core and non-cool core clusters, intermediate cases exist and play an important role in our understanding of cluster evolution (e.g., \citealt{2023A&A...670A.104M}). On the one hand, these intermediate cases can trace the transformation of a cool core into a non-cool core cluster, due to heating from major mergers or strong AGN feedback (e.g., \citealt{2010A&A...510A..83R,2023A&A...670A.104M,2025arXiv250313735G}). On the other hand, merger-induced turbulence and compression may also trigger localized cooling (e.g., \citealt{2013AN....334..394G,2015Natur.519..203V,2018ApJ...868..102V}). Therefore, these transitional clusters can provide insights into both the suppression and onset of cooling and feedback. 
\\\par A recent example is given by [SMG2021] CHIPS J1911+4455 (hereafter CHIPS~1911+4455). This recently discovered cluster at z=0.485 \citep{2021ApJ...907L..12S,2021ApJ...910...60S} defies the typical expectations of cool cores, as it not only harbors a starburst galaxy at its center (contrary to most cool cores, see \citealt{mcdonald2018}), but also exhibits a highly disturbed ICM morphology (Figure \ref{fig:multiwav}, left panel). Using 30 ks of Chandra X-ray observations, \citet{2021ApJ...907L..12S} revealed that CHIPS~1911+4455 has a strongly cooling core, with central entropy and cooling time among the lowest known (in the lowest 10\% of ACCEPT clusters, see \citealt{2009ApJS..182...12C}). Optical data from the {\it Hubble} Space Telescope (HST) and Nordic Optical Telescope (NOT) revealed a filamentary starburst BCG, with a huge star formation rate, SFR$\sim$140 -- 190~M$_{\odot}$/yr, compared to typical quenched BCGs (SFR$\leq$10~M$_{\odot}$/yr, e.g., \citealt{2022PhR...973....1D}). However, on larger scales, CHIPS~1911+4455 is one of the least symmetric cool core clusters known, with an asymmetry in the 7$^{\text{th}}$ percentile of the sample studied by \citet{2015MNRAS.449..199M}, indicative of a dynamically disturbed ICM. This is at odds with the known association of cool cores with relaxed
clusters and of non-cool cores with dynamically disturbed clusters (e.g., \citealt{burns2008,2011A&A...532A.123R,2023A&A...670A.104M}). The scenario proposed by \citet{2021ApJ...907L..12S} is that a major merger triggered local instabilities in the ICM, which started to rapidly cool and form stars. Support for this scenario comes from very recent Northern Extended Millimeter Array (NOEMA) observations of the CO(2-1) line in this BCG \citep{2025arXiv250420538C}. The observations uncovered molecular gas co-spatial with the BCG, with a total mass of about $1.9\times10^{11}$~M$_{\odot}$. The morphology and kinematics of the molecular gas were interpreted by \citet{2025arXiv250420538C} to reflect tidal interactions in the BCG related to the merger.
\\\par To date, no dedicated study has examined the state of the central AGN of this cool core cluster. Understanding this may reveal the fate of the cooling gas, which is believed to fuel both star formation and the SMBH in central galaxies (e.g., \citealt{2022PhR...973....1D} for a recent review). To this end, in this work we present new multi-frequency radio observations of the BCG in CHIPS~1911+4455. We describe the data in Section \ref{sec:data}, and present the results based on the radio images and the spatially resolved radio spectrum in Section \ref{sec:results}. We then discuss our results in Section \ref{sec:disc}, by comparing the new radio observations with the published X-ray/optical analysis, and report our conclusions in Section \ref{sec:concl}. 
\\\par We assume a flat $\Lambda$CDM cosmology with H$_0 =70$~km/s/Mpc and $\Omega_m=0.3$, giving a scale of $6$~kpc/" at z = 0.485. The spectral index $\alpha$ is defined as $S_{\nu} \propto \nu^{-\alpha}$, where $S_{\nu}$ is the flux density at frequency $\nu$.

\section{Observations and data reduction}\label{sec:data}
The data analyzed in this article consist of 8h of Very Long Baseline Array (VLBA) observations (4h at L and 4h at C band), and 10h of Jansky Very Large Array (JVLA) observations in the A configuration of the array (4h at P, 3h at L, and 3h at S band). A summary of the characteristics of these observations can be found in Table \ref{tab:data} of Appendix \ref{app:images}. 
\subsection{VLBA observations}\label{subsec:vlba}
The VLBA observations (project BU037, PI: Ubertosi) were performed on August 23, 2024 (C band, centered at 4.9~GHz), and on August 26, 2024 (L band, centered at 1.6 GHz). The source 3C345 was used as fringe finder, to determine instrumental delay and phases, and as bandpass calibrator, while the source J1921+4333, located at 2.3$^{\circ}$ from the target, was observed as phase reference calibrator. The VLBA data were reduced in AIPS using standard data reduction techniques\footnote{See \url{https://casaguides.nrao.edu/index.php/VLBA_AIPS_and_CASA_Walkthrough}.} and imaged in CASA 6.7.0 using a hogbom deconvolver with briggs weighting. We adopted a conservative 10\% uncertainty on the flux density scale of our observations. The C band data reached an rms noise of 23~$\mu$Jy/beam (beam FWHM 4.3$\times$1.4 mas, position angle $-1.4^{\circ}$), while the L band data reached an rms noise of 37~$\mu$Jy/beam (beam FWHM 11.3$\times$4.7 mas, position angle $-5.1^{\circ}$).

\begin{figure*}
    \centering
    \includegraphics[width=0.497\linewidth, trim={0 0 0.8cm 0}]{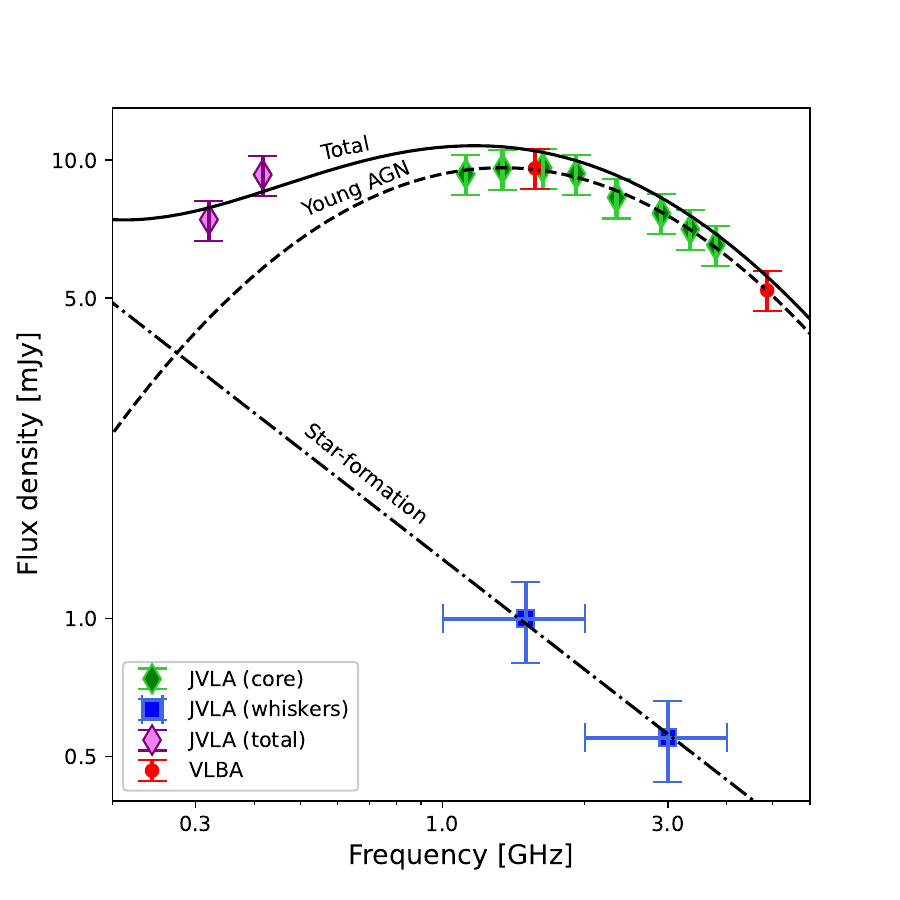}
    \includegraphics[width=0.497\linewidth,trim={0.8cm 0 0 0}]{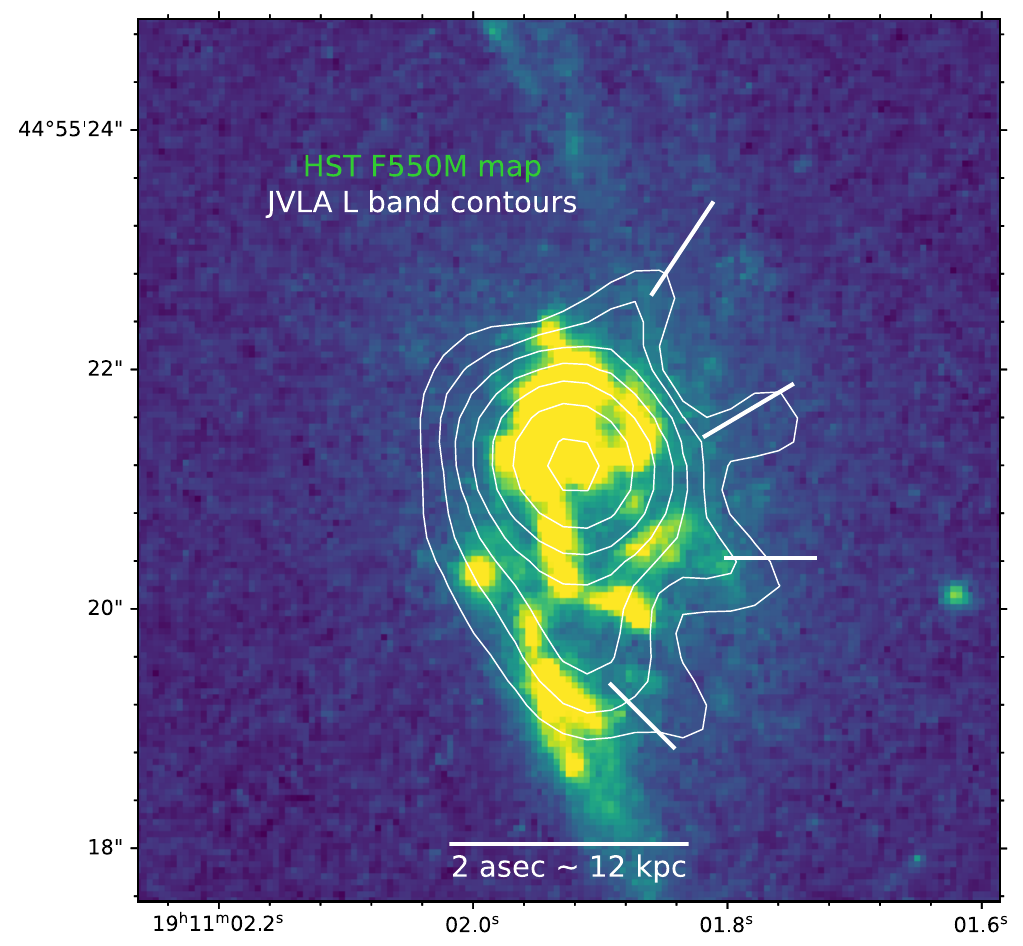}
    \caption{Radio synchrotron sources in CHIPS~1911+4455. \textit{Left:} JVLA and VLBA spectra of the radio core and the whiskers in CHIPS~1911+4455. Best-fit models to the spectra are overlaid as dashed-dotted (whiskers), dashed (radio core), and solid (total) lines (see Section \ref{subsec:sed}). \textit{Right:} HST F550M (sensitive to blue continuum and [O II] lines) image of the star-forming regions in the BCG with overlaid L-band contours (same as in Figure \ref{fig:vlba}).}
    \label{fig:sed}
\end{figure*}

\subsection{JVLA observations}\label{subsec:jvla}
The JVLA observations (project 24B-185, PI: Ubertosi) were performed on January 3, 2025 (S band, 2-4~GHz), January 4, 2025 (L band, 1-2 GHz), and January 18 and 23, 2025 (P band, 230-470~MHz). The source 3C~286 was used as primary calibrator (flux density scale, bandpass, delay, and phase), while the source J1845+4007, located at 6.2$^{\circ}$ from the target, was used as secondary calibrator (phase reference). The JVLA observations were reduced and imaged in CASA. The L band and S band data were reduced with the JVLA CASA pipeline\footnote{See \url{https://science.nrao.edu/facilities/vla/data-processing/pipeline/scripted-pipeline}.}, while the P band data was manually reduced in CASA v6.6.5 following standard data reduction techniques\footnote{See \url{https://casaguides.nrao.edu/index.php/VLA_Radio_galaxy_3C_129:_P-band_continuum_tutorial-CASA6.4.1}.}, as the pipeline does not handle data below 1~GHz. Extensive flagging in time and frequency of the calibrated P band data was necessary due the high percentage of the data affected by radio frequency interference (50\%) and due the observation being performed at sunrise. Furthermore, to mitigate the strong contamination from the Galactic plane (the target is located at a low Galactic latitude of $+15^{\circ}$), we excluded baselines smaller than 4$k\lambda$ in the uv-coverage, corresponding to angular scales larger than 50"$\sim$300~kpc (5 times the angular resolution). Overall, we flagged $\sim75\%$ of the data. 
We adopted a conservative 10\% uncertainty on the flux density scale of our observations. Our final broadband images were made with a multifrequency, multiscale approach in deconvolution. The resulting images reach rms noise levels near the source of 15~$\mu$Jy/beam for the L band (beam FWHM 0.9"$\times$0.8", position angle $38^{\circ}$), of 6~$\mu$Jy/beam for the S band (beam FWHM 0.4"$\times$0.4", position angle $-55^{\circ}$), and of 0.6~mJy/beam for the P band (beam FWHM 12"$\times$5", position angle $-60^{\circ}$).

\section{Results}\label{sec:results}
\subsection{Radio images and radio spectrum}\label{subsec:sed}
We report the VLBA detection of the BCG in CHIPS~1911+4455. The C band observation recovers a central point source with two-sided jets in the northeast - southwest direction (see Figure \ref{fig:multiwav}), each extending for about 5~mas (30~pc) from the core. The total flux density above 5$\times\sigma_{rms}$ at C band is of $S_{4.9} = 5.2\pm0.6$~mJy (peak flux 1.1~mJy/beam). The radio core is located at RA$=$19:11:01.9247, DEC$=$+44:55:21.244. In the L~band observation the source is unresolved, although hints of extensions east and west of the core are visible at 3$\times\sigma_{rms}$. The total flux density above 5$\times\sigma_{rms}$ is of $S_{1.6} = 9.6\pm0.9$~mJy. We thus derive an integrated spectral index of the radio source on parsec scales of $\alpha_{\text{VLBA}} = 0.55\pm0.08$. The VLBA images are shown in Figure \ref{fig:vlba} in Appendix \ref{app:images}. 
\\\par In the JVLA data at L band and S band, our images recover a central unresolved point source, surrounded by whiskers of extended radio emission to the south (best visible in the L band image), on scales of about 10~kpc (see Figure \ref{fig:multiwav}, top right panel, and Figure \ref{fig:vlba} in Appendix \ref{app:images}). The P band data, instead, show an unresolved point source due to the poorer angular resolution (Figure \ref{fig:vlba}). 
\\\par Figure \ref{fig:sed} (left panel) presents the broadband radio spectrum of the source, which we use to analyze the spectral properties of the BCG. We considered the total flux densities in the VLBA data at L band and C band (see Section \ref{subsec:vlba}), and the flux densities measured in JVLA data. For the JVLA data, in order to have a more discrete sampling of the spectrum, we measured flux densities in subbands at 320 MHz and 416 MHz (P band); at 1.12 GHz, 1.34 GHz, 1.63 GHz, 1.92 GHz (L band); and at 2.34 GHz, 2.9 GHz, 3.34 GHz, and 3.79 GHz (S band). As the target is unresolved in the P band data, the two low-frequency purple points in Figure \ref{fig:sed} (left) represent the total flux density. At L and S band, where the emission is resolved between the central point source and the whiskers, we measured the unresolved point source flux density in the above 8 subbands from Gaussian fits to the images (the green points in Figure \ref{fig:sed}, left). Instead, the flux density of the fainter southern whiskers was measured in the broadband L band and S band data to maximize the signal-to-noise ratio (blue points in Figure \ref{fig:sed}, left), by subtracting the point source flux density from the total flux density within the region encompassed by the 5$\times\sigma_{rms}$ level of the L band data. 
\\\par Based on the spectrum shown in Figure \ref{fig:sed} (left panel), the JVLA and VLBA flux density measurements are in excellent agreement, with the 1.6~GHz VLBA flux density corresponding to 99.7\% of the JVLA one. This match supports the idea that the parsec-scale radio emission from the BCG represents the maximum extent of the radio source, that is, the BCG radio jets extend for only 30~pc and are thus likely very young. Assuming a jet expansion speed of 0.1\,c (e.g., \citealt{2009AN....330..193G}), we can estimate a kinematic age of $\sim10^{3}$~yr. Overall, the JVLA flux densities at L and S band show a progressive flattening toward low frequencies, with a potential turnover below $\sim$1.5~GHz. This turnover is further supported by the 416~MHz flux density being 25\% higher than the 320~MHz one. The spectrum of the southern whiskers is described by a single powerlaw with spectral index $\alpha_{w} = 0.8\pm0.1$. Extrapolating this spectrum to the P band, we would expect a contribution of 2 - 4 mJy around 300 - 400 MHz. Therefore, both the unresolved radio core and the southern whiskers contribute to the total flux density at P band. To account for this in analyzing the JVLA data points of the unresolved core, we first subtracted the expected contribution of the southern whiskers from the integrated flux densities at P band. A curved spectrum model provides a good fit to the JVLA spectrum and returns a peak frequency of 1.3~GHz, which corresponds to $\nu_{p}^{rest} = 1.93$~GHz at z = 0.485. We also note that the flux density predicted by the model at 4.9~GHz ($5.16$~mJy) is in excellent agreement with the VLBA total flux density at the same frequency ($5.2\pm0.6$~mJy).

\section{Discussion}\label{sec:disc}
\subsection{AGN activity and star formation in the BCG}\label{subsec:interpret}
Our JVLA and VLBA data reveal that the radio emission from the core of the BCG is compact, unresolved on arcsec (kpc) scales, and shows the typical peaked spectrum of young radio sources. The spectral peak above 1~GHz, the active core and two-sided jets with largest linear size of about 60~pc in the VLBA data support the classification of the source as a GigaHertz Peaked Spectrum radio galaxy (e.g., \citealt{2009AN....330..193G,2016AN....337..105S}), that is, a SMBH whose activity as AGN started only very recently ($\sim10^{3}-10^{4}$~yr). 
\\\par The BCG is one of the most rapidly star-forming central cluster galaxies known in the z$\leq$1 Universe, with a SFR of $140 - 190$~M$_{\odot}$/yr \citep{2021ApJ...907L..12S}. As shown by narrowband HST images of the [OII] line at 3727\AA, the bulk of star formation is occurring in massive, filamentary knots extended to the south. We show in Figure \ref{fig:sed} (right) an overlay of the L band contours on the narrowband HST image. The extended radio whiskers coincide closely with the star-forming knots in the HST images, suggesting that the whiskers may represent star-formation-powered radio emission. At low ($\leq$30~GHz) frequencies, radio emission from star-forming regions is dominated by synchrotron emission due to supernovae (e.g., \citealt{1992ARA&A..30..575C,2018A&A...611A..55K}), with spectral index $\sim$0.8 around 1~GHz. Considering the relation between the SFR and 1.4~GHz radio luminosity, we can estimate the equivalent SFR of the southern whiskers. The 1.4 GHz flux density of the whiskers, 1.05~mJy (see Section \ref{subsec:sed}), 
corresponds to a luminosity of $L_{1.4} = 8.4\times10^{23}$~W/Hz. Depending on the adopted calibration of the $L_{1.4}$ -- SFR relation, the inferred SFR varies between  96~M$_{\odot}$/yr \citep{2017MNRAS.466.2312D}, 119~M$_{\odot}$/yr \citep{2024MNRAS.531..708C}, 155~M$_{\odot}$/yr \citep{1992ARA&A..30..575C}, and 120~M$_{\odot}$/yr \citep{2018MNRAS.475.3010G}, with the latter estimate based on a stellar mass of $2.2\times10^{11}$~M$_{\odot}$ \citep{2025arXiv250420538C}. This range of radio-derived SFR, 100 -- 155~M$_{\odot}$/yr, is comparable to the estimated SFR of 140 -- 190 M$_{\odot}$/yr from optical data \citep{2021ApJ...907L..12S}. Possible differences between the optical and radio SFRs may be influenced by the different regions from which the two estimates were derived, by the distinct timescales to which these indicators are sensitive, by the uncertain role of magnetic fields in amplifying the observed synchrotron flux density (e.g., \citealt{2013seg..book..419C}), and by uncertainties in the calibration of the $L_{1.4}$ -- SFR relation, which could also show a redshift dependence (e.g., \citealt{2021A&A...647A.123D}). We also acknowledge that, in our measurements of the radio SFR, we are neglecting contribution from star formation in the innermost 1", where the central young AGN is found. Based on the HST F550M image, we find that 15\% of the total [O~II] emission comes from this region. Increasing the radio SFR range of 100 -- 155 M$_{\odot}$/yr by 15\% we would obtain a new radio SFR of 115 -- 180 M$_{\odot}$/yr, that has a stronger overlap with the optical one. At the same time, we note that 
the 10\% relative uncertainty on the flux density scale of the JVLA data (Section \ref{subsec:jvla}) is comparable to the potential 15\% contribution from star formation discussed above. Overall, given all the above notes of caution, we argue that the radio and optical SFRs are broadly consistent.
\\ \par Another possible scenario is that the extended radio emission forming the whiskers is due to a collimated jet from the central AGN. However, there are a number of reasons that disfavor this scenario. Specifically, (1) the one-sidedness of the southern feature would imply that the radio galaxy is being seen nearly face-on (e.g., \citealt{1999MNRAS.306..513L}), which is not supported by the optical spectroscopy data in \cite{2021ApJ...907L..12S}, that do not see the typical AGN lines expected from a face-on active SMBH; (2) the morphology of the southern emission is strongly at odds with the expectations for a collimated jet, as it is composed of multiple extensions of comparable length in different directions (especially in the more sensitive L band data); (3) if a jet launched from the central AGN and extending toward the southern whiskers was present, the JVLA flux density of the unresolved core at 1~GHz should exceed the VLBA one; instead, the two datasets are in strong agreement (see Figure~\ref{fig:sed}, right, and Section~\ref{sec:results}); (4) the jets on pc scales would be misaligned by $\sim$70$^{\circ}$ from the putative kpc scale jet; while this is not impossible and other cases of pc-kpc scale mis-alignments are known (e.g., \citealt{1988ApJ...328..114P}), it adds non-negligible complexity to this scenario. 
\\\par We also disfavor the possibility that the extended synchrotron emission observed at L and S bands originates from relativistic plasma in a remnant, fossil lobe produced during a previous outburst and subsequently shaped by ICM motions relative to the BCG (as observed in other systems, e.g., \citealt{2022A&A...661A..92B}). In such a scenario, a steep spectral index would be expected due to synchrotron aging and adiabatic losses of the electron population ($\alpha\geq1.5$, e.g., \citealt{2001AJ....122.1172S}). However, the measured spectral index of the whiskers, $\alpha_{w} \sim 0.8$, is relatively flat and close to the injection spectrum of freshly accelerated relativistic electrons ($0.5\leq\alpha_{inj}\leq0.7$, e.g., \citealt{1973A&A....26..423J}).

\subsection{Clues on the onset of feedback in a dynamically unrelaxed cool core}
\indent Our new JVLA and VLBA data support a picture in which the BCG in CHIPS~1911+4455 has just turned on its radio jets. 
This BCG is known to inhabit a strong cool core \citep{2021ApJ...907L..12S}. In this sense, we note that CHIPS~1911+4455 resembles the ``pre-feedback'' clusters discussed in \cite{2023A&A...673A..52U}. These are cool cores where the central radio galaxy is young ($\sim10^{3}$~yr) and no evidence of recent kpc-scale AGN activity is present. Support for the recent onset of the central radio galaxy of these systems also comes from the X-rays: the AGN in pre-feedback clusters identified so far (see \citealt{2023A&A...673A..52U} for the initial selection, and \citealt{white2025} for another recent case) show relatively bright X-ray point sources, which supports the idea that these systems are actively accreting \citep{white2025}. The hot gas of these systems seems to reflect the unheated state of the cool core, with the entropy and cooling time in the innermost tens of kpc falling a factor $\sim$2 below those of ``mature-feedback'' clusters. Interestingly, {\it Chandra} data (30~ks, \citealt{2021ApJ...907L..12S}) showed that CHIPS~1911+4455 {\bf  has a hot} gas entropy at $\leq10$~kpc that is a factor $\sim$1.6 lower than the average of cool core ACCEPT \citep{2009ApJS..182...12C} clusters. Therefore, CHIPS~1911+4455 may represent another system, beyond the few identified in \cite{2023A&A...673A..52U}, where we can measure which hot gas properties are key in triggering the onset of AGN feedback. 
\\\par Notably, while low entropy and short cooling time of the hot gas appear to be common features of pre-feedback clusters known so far, the SFR of the central galaxy shows substantial variations -- ranging from $\geq$100~M$_{\odot}$/yr in CHIPS~1911+4455 (see \citealt{2021ApJ...907L..12S} and this work), to $\sim$10~M$_{\odot}$/yr in ClG~J0242-2132 and RXJ~1350.3+0940 (see \citealt{2023A&A...673A..52U}), to $\leq$1~M$_{\odot}$/yr in A1885 (see \citealt{white2025}). We will provide an extensive discussion of this apparent mismatch between different gas phases in our forthcoming paper on an extended selection of pre-feedback clusters (Ubertosi et al., in preparation). For now, we note that the data support a scenario in which the quenching of hot gas and star-forming gas proceeds on different timescales. This may suggest a partial decoupling between the long-term evolution of the hot and warm gas phases -- specifically, that the onset of jet feedback leads to the rapid suppression of hot gas cooling, but not necessarily of star formation (see also \citealt{2023A&A...673A..52U} and \citealt{white2025} for similar arguments).
\\ \par The fact that CHIPS~1911+4455 also shows large-scale ($\sim$100~kpc) significant asymmetries in the ICM, that \citet{2021ApJ...907L..12S} related to an ongoing merger, slightly complicates the picture of the conditions necessary to trigger the AGN activation. In the pre-feedback clusters from \cite{2023A&A...673A..52U}, the conclusion is that it is the lack of heating from AGN feedback for an extended period of time ($\geq$200~Myr) that triggers more rapid hot gas cooling. In CHIPS~1911+4455, the copious hot gas cooling may partially be caused by a merger-related increase of turbulence or compression. In any case, this system supports the idea that about twice lower entropy of the hot gas is a key condition in determining the onset of mechanical AGN feedback. 
\\ \par We also note that systems such as CHIPS~1911+4455 might be more common at redshifts $z\geq0.5$, where unrelaxed clusters tend to host more strongly star-forming central galaxies (e.g., \citealt{mcdonald2016}). High resolution radio follow-up of these high-redshift systems might reveal more analogous cases of unrelaxed galaxy clusters with strong cool cores and newly-activated central SMBHs.

\section{Conclusion}\label{sec:concl}
In this article, we presented new VLBA and JVLA observations of the starburst BCG in CHIPS~1911+4455, a peculiar cool core cluster at z = 0.485 with merger signatures. Our multifrequency (320 MHz -- 5 GHz) and multiscale (0.01 -- 20 kpc) analysis of this dataset reveals that the central radio galaxy in the BCG has recently switched on. The AGN appears to be at a very early stage in its lifecycle, being composed of a compact core with two-sided jets each extending for 30~pc. At larger scales, our JVLA L and S band observations detect radio emission extending over $\sim$10 kpc, forming a set of whiskers south of the BCG.  These show a striking spatial match to the star-forming knots of this starburst BCG previously identified in HST images, and are thus most likely powered by synchrotron emission. The radio luminosity of the whiskers returns a SFR of 100 -- 155~M$_{\odot}$/yr, close to the optical/infrared estimate of 140 -- 190~M$_{\odot}$/yr. We propose that the AGN activation in this system is related to a lower ICM entropy (nearly a factor of 2) around the BCG, tracing efficient cooling, with respect to systems with full-fledged central radio galaxies. Overall, CHIPS~1911+4455 may represent a different ``flavor'' of pre-feedback clusters (see \citealt{2023A&A...673A..52U}), in which the extensive cooling that triggers AGN feedback is not only the consequence of a prolonged pause in the SMBH activity, but also of the ongoing dynamical disturbances.


\begin{acknowledgements}
      We thank the reviewer for their constructive and useful suggestions on our work. FU thanks M. Balboni for useful suggestions on low Galactic latitude radio observations. FU and MG acknowledge support from the research project PRIN 2022 ``AGN-sCAN: zooming-in on the AGN-galaxy connection since the cosmic noon", contract 2022JZJBHM\_002 -- CUP J53D23001610006. PT acknowledges support from NASA’s NNH22ZDA001N Astrophysics Data and Analysis Program under award 24-ADAP24-0011. VO acknowledges support from the research projects DYCIT ESO-Chile Comite Mixto PS 1757, Fondecyt Regular 1251702, and DICYT PS 541. The National Radio Astronomy Observatory is a facility of the National Science Foundation operated under cooperative agreement by Associated Universities, Inc. Some of the data presented in this paper were obtained from the Mikulski Archive for Space Telescopes (MAST) at the Space Telescope Science Institute. The specific observations analyzed can be accessed via \dataset[https://doi.org/10.17909/ppck-k333]{https://doi.org/10.17909/ppck-k333}. STScI is operated by the Association of Universities for Research in Astronomy, Inc., under NASA contract NAS5–26555. Support to MAST for these data is provided by the NASA Office of Space Science via grant NAG5–7584 and by other grants and contracts.
\end{acknowledgements}

\begin{contribution}
FU developed the research concept, analyzed the data, and was responsible for writing and submitting the manuscript. 
All the co-authors contributed equally to reviewing the manuscript.   
\end{contribution}

\facilities{NRAO, CXO, HST}




\begin{appendix}
\section{JVLA and VLBA radio data and images}\label{app:images}
\begin{figure*}[ht!]
    \centering
    \includegraphics[width=\linewidth]{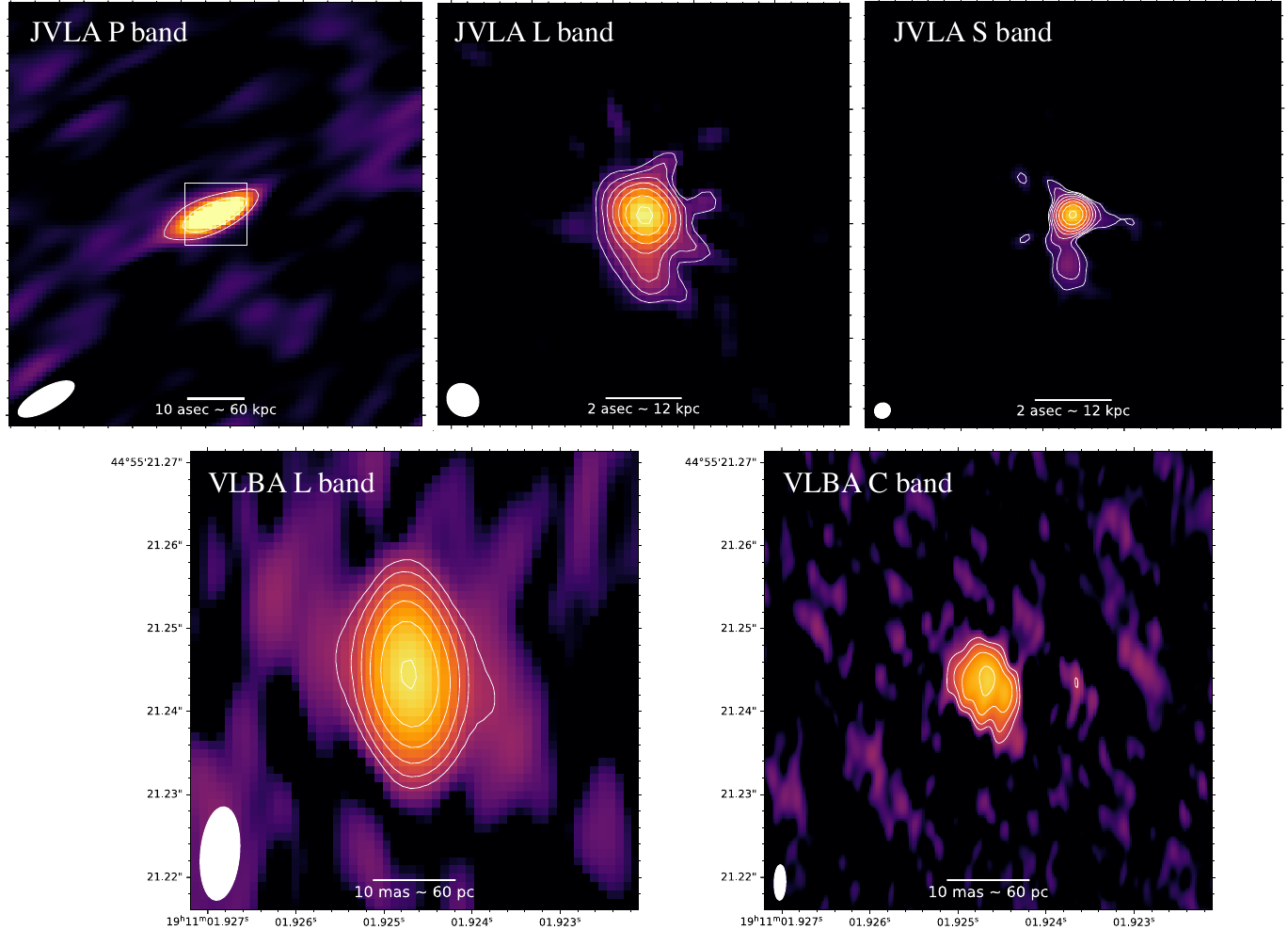}
    \caption{New radio images of the BCG in CHIPS~1911+4455. \textit{Top left:} JVLA image at P band (centered at 370~MHz; rms noise of 0.6~mJy/beam, beam FWHM 12"$\times$5", position angle $-60^{\circ}$), with a white square showing the field of view of the JVLA L band and S band images shown in the top center and right panels; \textit{Top center:} JVLA image at L band (centered at 1.5 GHz; rms noise of 15~$\mu$Jy/beam, beam FWHM 0.9"$\times$0.8", position angle $38^{\circ}$); \textit{Top right:} JVLA image at S band (centered at 3.0 GHz; rms noise of 6~$\mu$Jy/beam, beam FWHM 0.4"$\times$0.4", position angle $-55^{\circ}$) JVLA image; \textit{Bottom left:} VLBA image at L band (centered at 1.6 GHz; rms noise of 37~$\mu$Jy/beam, beam FWHM 11.3$\times$4.7 mas, position angle $-5.1^{\circ}$); \textit{Bottom right:} VLBA image at C band (centered at 4.9 GHz; rms noise of 23~$\mu$Jy/beam, beam FWHM 4.3$\times$1.4 mas, position angle $-1.4^{\circ}$). Contours start at 5$\times$ the rms noise close to the source and increase by a factor of 2. The beam is shown with a white ellipse in the bottom left corner.}\label{fig:vlba}
\end{figure*}

We show in Figure~\ref{fig:vlba} the JVLA broadband images at P band, L band, and S band, and the VLBA images at L band and at C band. A summary of the radio observations and images is reported in Table \ref{tab:data}. The description of the data and the reduction techniques are reported in Section~\ref{sec:data}.

 \begin{table}[]
     \centering
     \caption{Summary of radio observations employed in this work and properties of the images shown in Fig.~\ref{fig:vlba}. (1) Observing band and central frequency; (2) date of the observations; (3) total time; (4) rms noise of the images; (5) angular resolution; (6) peak flux density; (7) total flux density above 5$\times\sigma_{rms}$ .}\label{tab:data}\renewcommand*{\arraystretch}{1.2}
     \begin{tabular}{c|c|c|c|c|c|c}
\hline
 \multicolumn{6}{c}{{ VLBA observations (Project BU037) }} \\
\hline
 Obs. band & Obs. Date & t$_{obs}$ & $\sigma_{rms}$ & Beam FWHM & S$_{p}$ & S$_{tot}$ \\
 \hline
 
  L band (1.6 GHz) & Aug. 26, 2024 & 4h & 37~$\mu$Jy/beam & 11.3$\times$4.7~mas, $-5.1^{\circ}$ & $6.3$~mJy/beam & $9.6\pm0.9$~mJy\\
 
  C band (4.9 GHz) & Aug. 23, 2024 & 4h & 23~$\mu$Jy/beam & 4.3$\times$1.4~mas, $-1.4^{\circ}$ & $1.1$~mJy/beam & $5.2\pm0.6$~mJy\\

\hline

 \multicolumn{6}{c}{{ JVLA observations (Project 24B-185)}} \\

 \hline

 P band (370 MHz) & Jan. 18 \& 25, 2025 & 4h & 600~$\mu$Jy/beam & 12"$\times$5", $-60^{\circ}$ & $9.0$~mJy/beam & $8.4\pm0.8$~mJy\\

 L band (1.5 GHz) & Jan. 4, 2025 & 3h & 15~$\mu$Jy/beam & 0.9"$\times$0.8", $38^{\circ}$ & $8.0$~mJy/beam & $9.5\pm1.0$~mJy\\
 
 S band (3.0 GHz) & Jan. 3, 2025 & 3h & 6~$\mu$Jy/beam & 0.4"$\times$0.4", $-55^{\circ}$ & $6.4$~mJy/beam & $7.3\pm0.7$~mJy\\
 
 \hline

  \hline

\end{tabular}

 \end{table}

\end{appendix}

\bibliographystyle{aasjournal}
\bibliography{bibliography}{} 
\end{document}